\documentclass[aps,prd,,twocolumn,superscriptaddress]{revtex4-1}

\usepackage{graphicx}
\usepackage{amsmath}

\newcommand{\Lagr}{\mathcal{L}}
\newcommand{\pmass}{m_{\gamma\prime}}

\begin{document}

\title{Hidden Sector Photon Coupling of Resonant Cavities}

\author{Stephen R. Parker}
\email{stephen.parker@uwa.edu.au}
\affiliation{School of Physics, The University of Western Australia, Crawley 6009, Australia}
\author{Gray Rybka}
\affiliation{University of Washington, Seattle, Washington 98195, USA}
\author{Michael E. Tobar}
\affiliation{School of Physics, The University of Western Australia, Crawley 6009, Australia}

\date{\today}

\begin{abstract}
Many beyond the standard model theories introduce light paraphotons, a hypothetical spin-1 field that kinetically mixes with photons. Microwave cavity experiments have traditionally searched for paraphotons via transmission of power from an actively driven cavity to a passive receiver cavity, with the two cavities separated by a barrier that is impenetrable to photons. We extend this measurement technique to account for two-way coupling between the cavities and show that the presence of a paraphoton field can alter the resonant frequencies of the coupled cavity pair. We propose an experiment that exploits this effect and uses measurements of a cavities resonant frequency to constrain the paraphoton-photon mixing parameter, $\chi$. We show that such an experiment can improve sensitivity to $\chi$ over existing experiments for paraphoton masses less than the resonant frequency of the cavity, and eliminate some of the most common systematics for resonant cavity experiments.
\end{abstract}


\maketitle

\section{Introduction}

Some Standard Model extension theories postulate the existence of a hidden sector of particles that interact very weakly with standard model particles~\cite{sme1,sme2}. One proposed form of hidden sector particle interaction is the spontaneous kinetic mixing of the photon and the hidden sector photon~\cite{holdem}. Massive hidden sector photons are known as paraphotons~\cite{paraphoton} and are classified as a type of Weakly Interacting Slim Particle (WISP), a hypothetical group of particles with sub-eV masses~\cite{wisps}.  Experiments to detect paraphotons place bounds on the kinetic mixing parameter, $\chi$, as a function of the possible hidden sector photon mass.

Laboratory based searches for paraphotons have been conducted for several years~\cite{exp1,exp2,exp3,exp4,exp5,exp6,exp7} with some recent tests using microwave frequency resonant cavities~\cite{povey2010,ADMX2010,povey2011}. Electromagnetic resonances in otherwise isolated cavities could become coupled in the presence of a paraphoton field.  If one resonant cavity is actively driven, this coupling can be seen as photons in the driven cavity mixing with paraphotons, which then cross the boundary between cavities, and then mixing back into photons in the undriven cavity. Resonant regeneration is present even at the subphoton level~\cite{hartnett2011}, and by measuring the power transmitted between the two cavities a bound can be placed on the probability of kinetic mixing between photons and paraphotons. This arrangement is known as a Light Shining through a Wall (LSW) experiment and has been the focus of microwave frequency resonant cavity paraphoton searches~\cite{povey2010,ADMX2010}. As these searches rely on measuring very low levels of microwave power the fundamental limitation to their sensitivity is imposed by the thermal noise in the detector cavity and amplification system. However, in practice they have been limited by microwave power leakage from the emitter to detector cavity which is indistinguishable from a paraphoton effect~\cite{povey2010}. The prospect of LSW has also inspired some speculative work on exploiting the paraphoton for data transmission and communications~\cite{alpcomms1,alpcomms2}.

Previous LSW formalism~\cite{Jaeckel08} has been focused on the one way flow of paraphotons from a driven emitter cavity to an undriven detection cavity. However, it is also possible to treat the two-way exchange of paraphotons as a weak coupling between the cavities, creating a system analogous to two spring-mass oscillators connected via a third weak spring. When both cavities are actively driven the paraphoton mediated coupling will cause a phase-dependent shift in the resonant frequencies and quality factors of the system. This opens up the possibility of conducting experiments that constrain the strength of photon-paraphoton mixing by observing this coupling induced resonant frequency shift. When given the option it is preferable to make a measurement of frequency rather than power due to the quality and precision of frequency standards, instrumentation and techniques.

Although we focus on the paraphoton, the concepts developed in this paper can be extended and applied to LSW based searches for other hypothetical particles that mix with the photon such as fermionic minicharged particles~\cite{MCPLSW1,MCPLSW2}.

\section{Fundamental Normal Modes}

As described in Jaeckel $\&$ Ringwald~\cite{Jaeckel08} the renormalizable Lagrangian for low energy photons and paraphotons is given by
\begin{equation}
\begin{split}
\Lagr = -\frac{1}{4}F^{\mu\nu}F_{\mu\nu} - \frac{1}{4}B^{\mu\nu}B_{\mu\nu} \\
- \frac{1}{2}\chi F^{\mu\nu} B_{\mu\nu} + \frac{1}{2} m^{2}_{\gamma\prime}B_{\mu}B_{\mu},
\label{eq:lagrangian}
\end{split}
\end{equation}
where $F^{\mu\nu}$ is the field strength tensor for the photon field $A^{\mu}$, $B^{\nu\mu}$ is the field strength tensor for the paraphoton field $B^{\mu}$, $\chi$ is the photon-paraphoton kinetic mixing parameter and $m_{\gamma\prime}$ is the paraphoton mass. From eq.~\eqref{eq:lagrangian} the equations of motion for the electromagnetic fields in two spatially separated resonant cavities, $A_{1}$ and $A_{2}$, and the universal paraphoton field $B$ are
\begin{align}
\left(\partial^{\mu}\partial_{\nu} + \chi^{2} m^{2}_{\gamma\prime}\right)A_{1} = \chi m^{2}_{\gamma\prime} B \label{eq:eoma1} \\
\left(\partial^{\mu}\partial_{\nu} + \chi^{2} m^{2}_{\gamma\prime}\right)A_{2} = \chi m^{2}_{\gamma\prime} B \label{eq:eoma2} \\
\left(\partial^{\mu}\partial_{\nu} + m^{2}_{\gamma\prime}\right)B = \chi m^{2}_{\gamma\prime}\left(A_{1}+A_{2}\right). \label{eq:eomb}
\end{align}
The cavity fields can be broken down in to time and spatial components,
\begin{equation}
A_{1,2}\left(\mathbf{x},t\right)=a_{1,2}\left(t\right)A_{1,2}\left(\mathbf{x}\right),
\end{equation}
where the spatial component satisfies the normalisation condition
\begin{equation}
\int\limits_V d^{3}\mathbf{x}|A_{1,2}\left(\mathbf{x}\right)|^{2}=1. \label{eq:norm}
\end{equation}
\begin{figure}[t!]
\centering
\includegraphics[width=0.95\columnwidth]{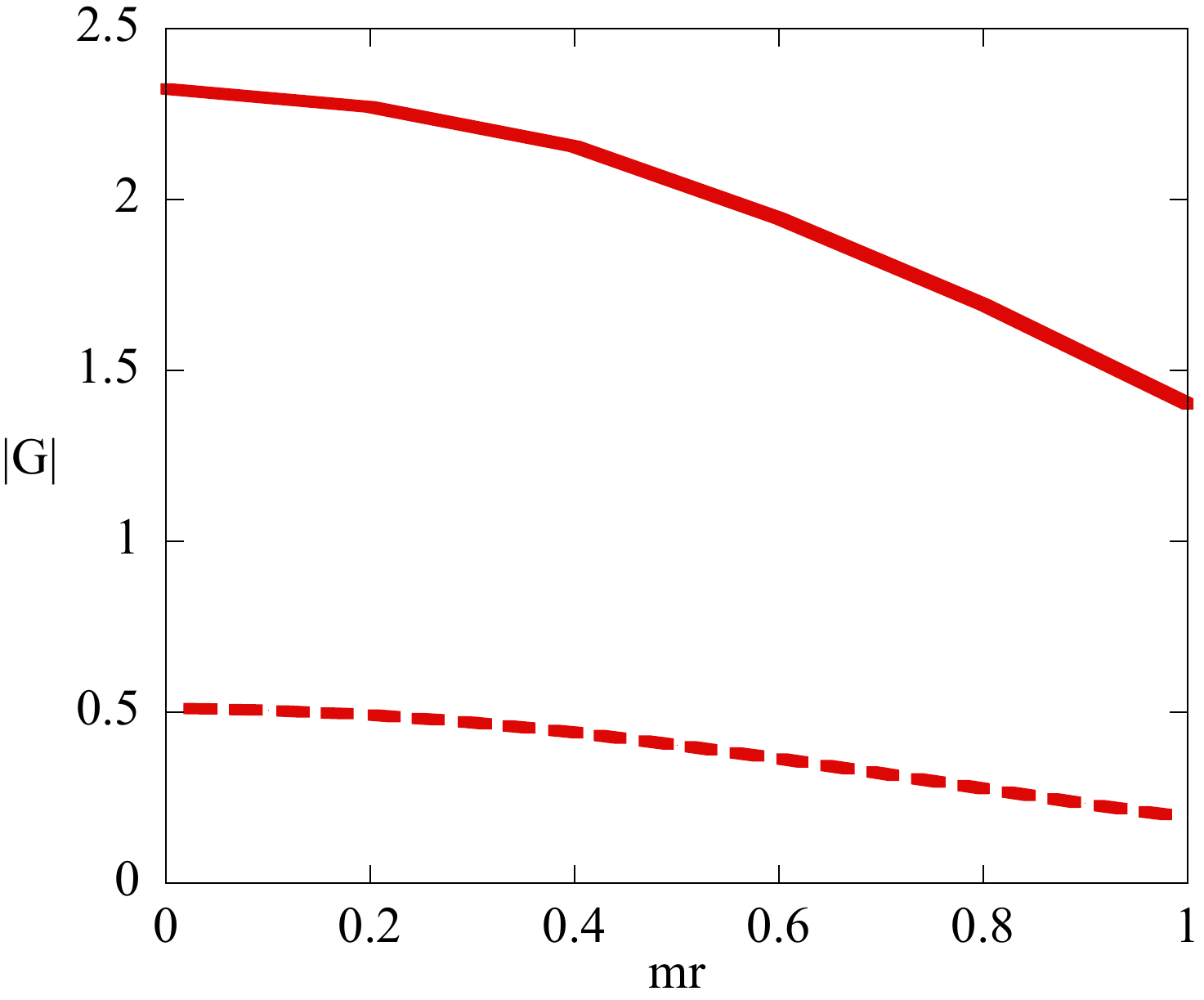}
\caption{(color online) Numerical simulation of the absolute value of geometry factors as a function of the paraphoton mass ratio, $\text{mr}=\pmass/\omega_{0}$, for two identical cube cavities overlapping in space ($G_{S}$, full line) and separated by a distance of L ($G$, dashed line) where L is the length of the cubic cavities, in this case set to 1. The normalized resonant mode used is given by $A(\mathbf{x})=2\sin{(\pi x)}\sin{(\pi y)}$ and the resonant frequency is $\omega_{0}=\sqrt{2}\pi$~\cite{Jaeckel08}.}
\label{fig:Gfacs}
\end{figure}
Due to the infinite nature of the paraphoton field we use the retarded massive Greens function to find the paraphoton field from equation~\eqref{eq:eomb},
\begin{equation}
\begin{split}
B\left(\mathbf{x},t\right)=\chi\pmass^{2}\left(\int\limits_{V1} d^{3}\mathbf{y}\frac{\text{exp}\left(ik_{b}|\mathbf{x}-\mathbf{y}|\right)}{4\pi|\mathbf{x}-\mathbf{y}|}a_{1}A_{1}\left(\mathbf{y}\right) \right. \\
\left. +\int\limits_{V2} d^{3}\mathbf{y}\frac{\text{exp}\left(ik_{b}|\mathbf{x}-\mathbf{y}|\right)}{4\pi|\mathbf{x}-\mathbf{y}|}a_{2}A_{2}\left(\mathbf{y}\right)\right). \label{eq:parafield}
\end{split}
\end{equation}
In the absence of a paraphoton field the resonant frequency, $\omega_{0}$, of a cavity is given by
\begin{equation}
\nabla^{2}A\left(\mathbf{x}\right)=\omega_{0}^{2}A\left(\mathbf{x}\right).
\label{eq:8}
\end{equation}
\begin{figure}[t!]
\centering
\includegraphics[width=0.98\columnwidth]{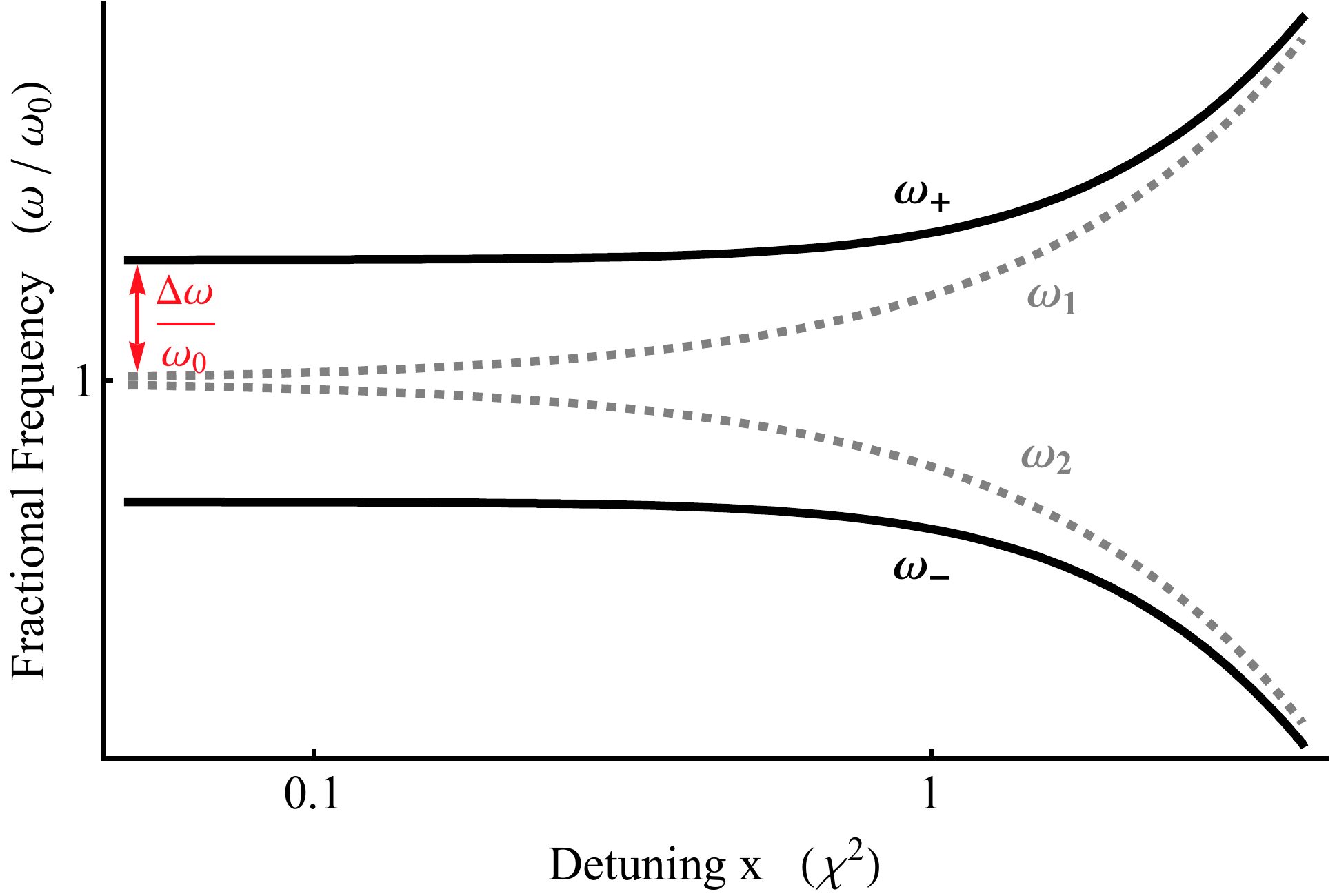}
\caption{(color online) Log-Log plot of resonant frequencies relative to a common central frequency, $\omega_{0}$, as a function of detuning for a pair of cavities that are coupled (black, full, equation~\eqref{eq:freqcoupled}) and uncoupled (gray, dashed, $\omega_{1,2} / \omega_{0} = \left(1\pm x / 2\right)$). The detuning, x, is given as a factor of the square of the paraphoton kinetic mixing parameter, $\chi$. The magnitude of the fractional frequency shift is proportional to the values of parameters $Q_{1}$, $Q_{2}$, $G$, $G_{S}$ and $\chi$ used in eq.~\eqref{eq:freqcoupled}.}
\label{fig:coupledfreqs}
\end{figure}
We can now solve for the photon field in cavity 1 by substituting the paraphoton field of eq.~\eqref{eq:parafield} in to eq.~\eqref{eq:eoma1} and utilizing eq.~\eqref{eq:8} and the normalization conditions from~\eqref{eq:norm} we find that:
\begin{align}
\begin{split}
\left(\omega_{0}^{2}-\omega_{1}^2-i\frac{\omega_{0}\omega_{1}}{Q_{1}}+\chi^{2}\pmass^{2}\left(1-\frac{\pmass^{2}}{\omega_{0}^{2}}G_{11}\right)\right)a_{1}\left(t\right) \\
=\frac{\chi^{2}\pmass^{4}G_{12}}{\omega_{0}^{2}}a_{2}\left(t\right) \label{eq:field1}
\end{split} \\
\begin{split}
G_{11}=\omega_{0}^{2}\int\limits_{V1}d^{3}\mathbf{x}\int\limits_{V1}d^{3}\mathbf{y}\frac{\text{exp}\left(ik_{b}|\mathbf{x}-\mathbf{y}|\right)}{4\pi |\mathbf{x}-\mathbf{y}|} \\
\times A_{1}\left(\mathbf{y}\right)\cdot A_{1}\left(\mathbf{x}\right)
\end{split} \nonumber \\
\begin{split}
G_{12}=\omega_{0}^{2}\int\limits_{V1}d^{3}\mathbf{x}\int\limits_{V2}d^{3}\mathbf{y}\frac{\text{exp}\left(ik_{b}|\mathbf{x}-\mathbf{y}|\right)}{4\pi |\mathbf{x}-\mathbf{y}|} \\
\times A_{2}\left(\mathbf{y}\right)\cdot A_{1}\left(\mathbf{x}\right), \label{eq:gfac}
\end{split}
\end{align}
where $\omega_{1}$ is the driving frequency of the cavity and $G_{12}$ is the standard G-function found in the literature~\cite{Jaeckel08,povey2010} that describes the two cavity fields, geometries and relative positions while $G_{11}$ (henceforth $G_{S}$) is the G-function for a cavity field overlapped spatially with itself and represents losses in the cavity due to photon to paraphoton conversion. Numerical calculations of $G$ and $G_{S}$ for a simple cube cavity system (Fig.~\ref{fig:Gfacs}) indicate that the magnitude of $G_{S}$ is larger than the magnitude of $G$ as one would expect. We can represent the total electrical cavity loss as a new quality factor,
\begin{equation}
Q_{T}=\left(\frac{1}{Q}+\chi^{2}\text{Im}\left(G_{S}\right)\right)^{-1},
\end{equation}
where classical electrical losses in the cavity due to surface resistance are represented by the traditional finite quality factor, $Q$. In principle cavity loss measurements could be used to set limits on the value of $\chi$, although this is not a practical or realistic way to constrain the kinetic mixing of the paraphoton. 

Following the method outlined earlier we can solve for the photon field in cavity 2 and hence relate the fields in the two cavities without using the paraphoton field, $B\left(\mathbf{x},t\right)$. This set of coupled equations can be represented in matrix form as
\begin{widetext}
\begin{equation}
\begin{bmatrix}
\omega_{0}^{2} - \omega_{1}^{2} - \frac{i \omega_{0}\omega_{1}}{Q_{1}}+\chi^{2}\pmass^{2}\left(1-\frac{\pmass^{2}}{\omega_{0}^{2}}G_{S}\right) & -\frac{\chi^{2}\pmass^{4}G}{\omega_{0}^{2}} \\
-\frac{\chi^{2}\pmass^{4}G}{\omega_{0}^{2}} & \omega_{0}^{2} - \omega_{2}^{2} - \frac{i \omega_{0}\omega_{2}}{Q_{2}}+\chi^{2}\pmass^{2}\left(1-\frac{\pmass^{2}}{\omega_{0}^{2}}G_{S}\right)
\end{bmatrix}.
\begin{bmatrix}
a_{1}(t) \\
a_{2}(t)
\end{bmatrix}=
\begin{bmatrix}
0 \\
0
\end{bmatrix},
\label{eq:coupledmatrix}
\end{equation}
\end{widetext}
with $G_{11}=G_{22}=G_{S}$ and $G_{12}=G_{21}=G$. In an ideal situation both cavities would be driven at the same frequency, $\omega$, however in reality their resonant frequencies are likely to differ by a small amount. We parameterize this detuning by the factor $x$ such that $\omega_{1}=\omega\left(1+\frac{x}{2}\right)$ and $\omega_{2}=\omega\left(1-\frac{x}{2}\right)$. We can find the two fundamental normal mode frequencies of the coupled cavity system by taking the determinant of eq.~\eqref{eq:coupledmatrix}, equating the real components to zero and solving for $\omega$, yielding
\begin{equation}
\begin{split}
\omega_{\pm}\approx\omega_{0}\left(\frac{1}{1-\frac{x^2}{2}}\left(1+\frac{1}{2Q_{1}Q_{2}}+\frac{x^2}{4}+\frac{\pmass^{2}\chi^{2}}{\omega_{0}^{2}} \right.\right. \\
\left.\left. -\frac{\pmass^{4}\chi^{2}G_{S}}{\omega_{0}^{4}} \pm\left(\frac{1}{Q_{1}Q_{2}}+x^{2}+\frac{2\pmass^{2}x^{2}\chi^{2}}{\omega_{0}^{2}} \right.\right.\right. \\
\left.\left.\left.-\frac{2\pmass^{4}x^{2}\chi^{2}G_{S}}{\omega_{0}^{4}}+\frac{\pmass^{8}\chi^{4}G}{\omega_{0}^{8}}\right)^{\frac{1}{2}}\right)\right)^{\frac{1}{2}},
\label{eq:freqcoupled}
\end{split}
\end{equation}
where some insignificant higher order terms have been removed. The effect of frequency detuning is demonstrated in figure~\ref{fig:coupledfreqs}; as the cavities become detuned the strength of the coupling weakens and the resonant frequencies approach their uncoupled values. The fractional frequency shift for a cavity due to the paraphoton coupling is illustrated in red, this is the value that any experiment would seek to measure. 

The coupled resonant modes associated with equation~\eqref{eq:freqcoupled} will also have different quality factors (assuming that the initial uncoupled cavity mode quality factors $Q_1$ and $Q_2$ are not identical). Exploiting this effect appears to offer no advantages over existing power based measurements such as LSW~\cite{povey2010,ADMX2010} and threshold crossing~\cite{povey2011} experiments, as such we shall focus our attention on frequency effects.

\section{Rotating Cavity Experiment}

\begin{figure}[t]
\centering
\includegraphics[width=0.85\columnwidth]{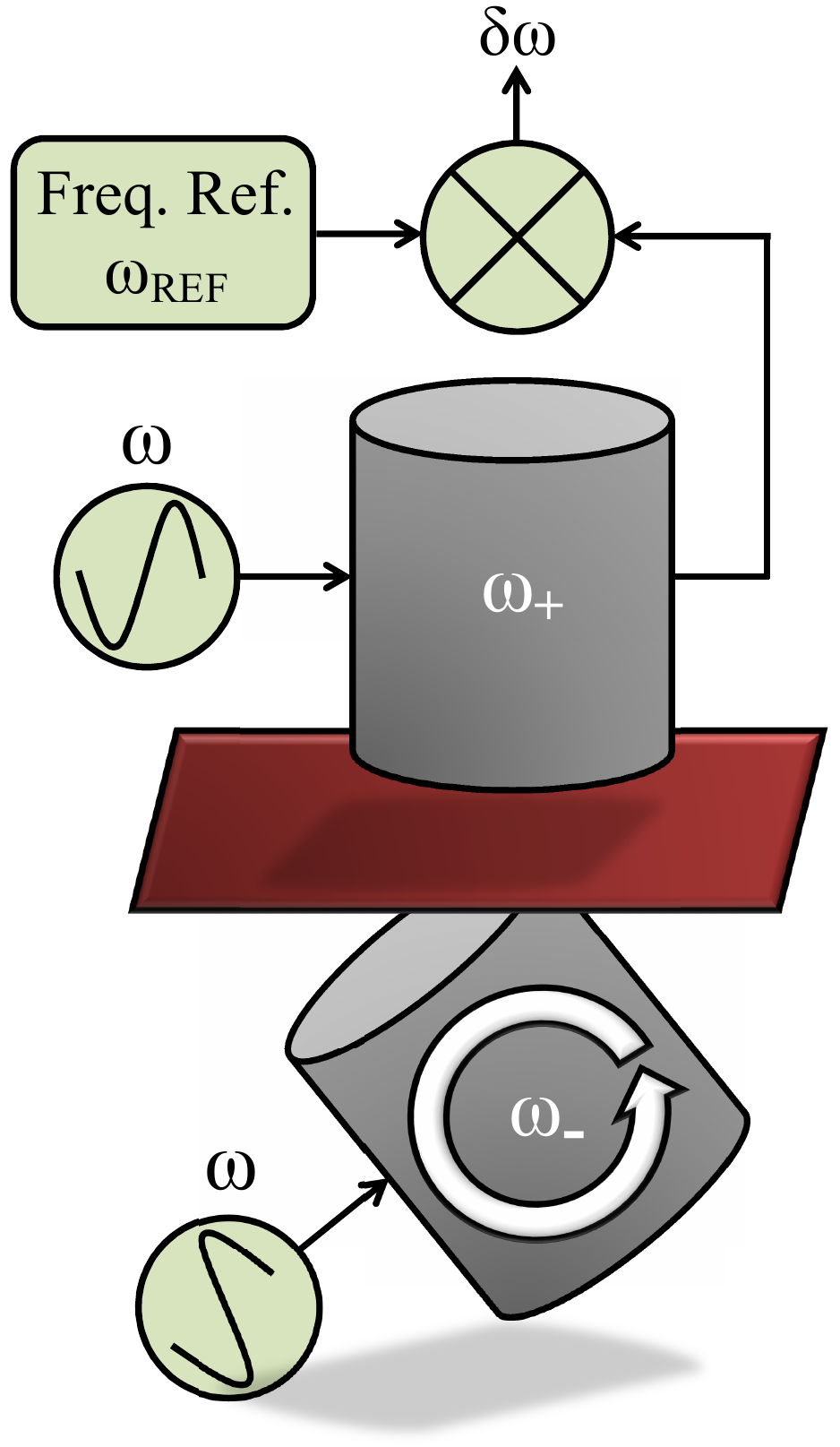}
\caption{(color online) Schematic of a rotating paraphoton coupled cavities experiment. The two empty cavities separated by an impenetrable barrier (red plane) and actively driven at frequency $\omega$, although the resonant frequencies of the cavities may be slightly detuned ($\omega_{+}$ and $\omega_{-}$). One cavity is rotated orthogonally to the other. The frequency of the stationary cavity is compared to a stable frequency reference ($\omega_{REF}$) to produce the beat frequency ($\delta\omega$).}
\label{fig:experiment}
\end{figure}

One of the most effective ways to measure the normal mode frequency shift would be to modulate the strength of the coupling between the cavities and look for the induced modulated signal in the beat frequency of a coupled cavity and an uncoupled frequency reference. This allows for a fast rate of data collection and reduces the influence of longterm frequency drift. The strength of the coupling between the cavities can be changed by manipulating the value of the G-factor from equation~\eqref{eq:gfac}. Changes to the relative alignment and separation of the two cavities will in turn alter the dot product of the two cavity fields and hence the G-factor. If one cavity is rotated orthogonally to the other then the dot product of the photon fields will be modulated sinusoidally at twice the rotation frequency, giving a maximum and minimum G-factor every half rotation. Rotating microwave cavities have been used in experiments to test for violations of local Lorentz invariance~\cite{LI05} and this approach has the added benefit of suppressing systematic signal leakage between the cavities to second order as the power leakage is modulated at a different frequency. As shown in figure~\ref{fig:experiment} the frequency of the stationary cavity can be compared against a frequency reference which must exhibit a higher frequency stability than the coupled cavities. Paraphoton mediated coupling between the test cavity and the frequency reference would be suppressed due to the relatively large separation distance. An ideal candidate for the reference frequency would be a Cryogenic Sapphire Oscillator (CSO) that is capable of generating a microwave signal with a fractional frequency stability of parts in $10^{-16}$ for integration times up to at least 100 seconds~\cite{CSO2008,CSO2012}. A CSO also has low level of paraphoton production as the majority of the electric field in the resonator is confined within the sapphire dielectric.

\begin{figure}[t!]
\centering
\includegraphics[width=0.98\columnwidth]{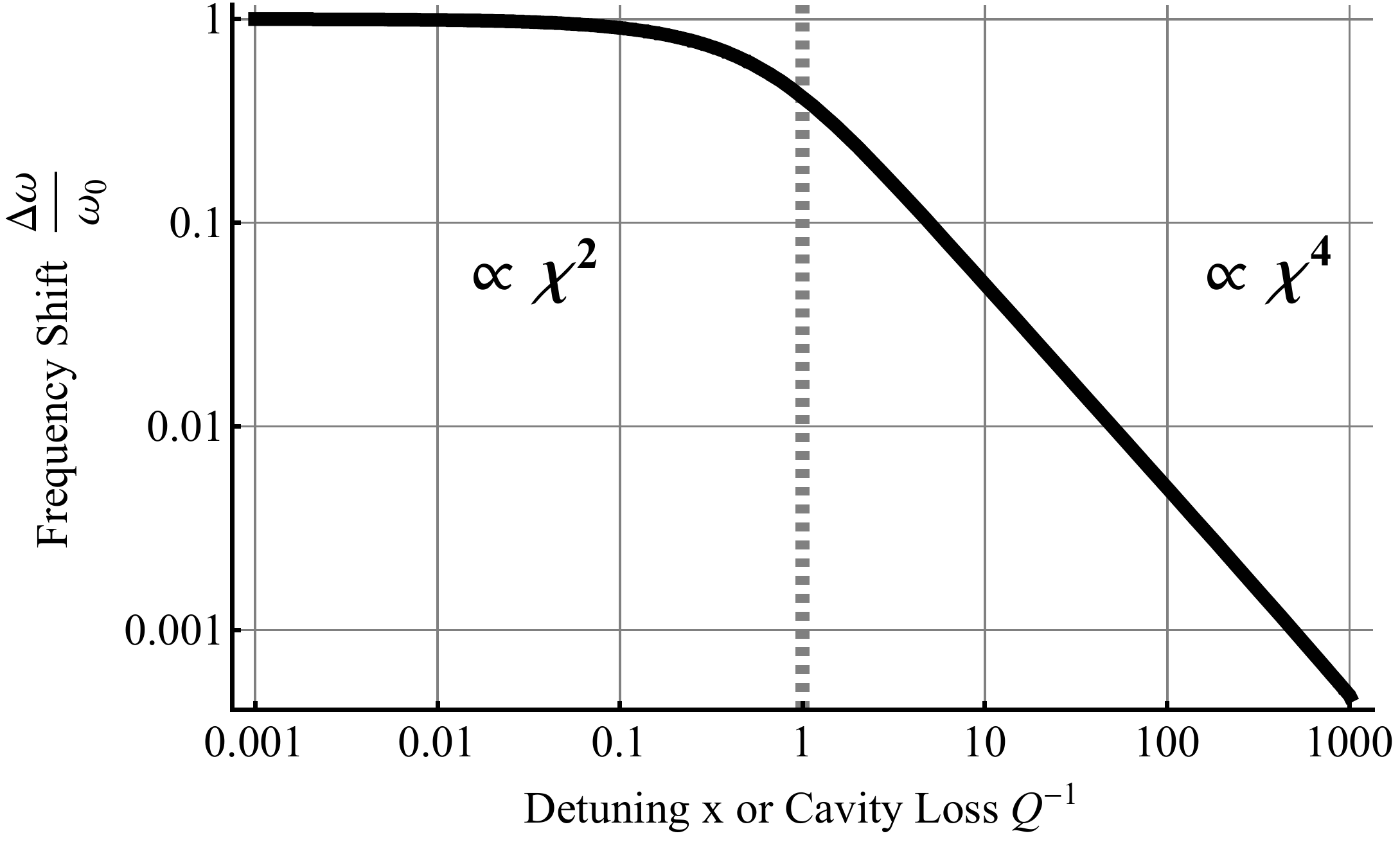}
\caption{(color online) Log-Log plot of the normalized fractional resonant frequency shift due to paraphoton cavity coupling as a function of cavity frequency detuning and losses (from eq.~\eqref{eq:freqcoupled} and eq.~\eqref{eq:deltaomega}). Both the cavity frequency detuning, $x$, and the cavity losses, $Q^{-1}$, are given as a factor of $\chi^{2}$. The frequency shift is plotted a function of $x$ (with $Q^{-1}=0$) and as a function of $Q^{-1}$ (with $x=0$); both curves overlap. The vertical gray dashed line indicates where the dependence of the frequency shift changes. In the first region the frequency shift is proportional to $\chi^{2}$ and in the second region it is proportional to $\chi^{4}$.}
\label{fig:freqshift}
\end{figure}

To determine the sensitivity of this experiment we need an expression that links the value of the paraphoton kinetic mixing parameter $\chi$ to the beat frequency of the stationary coupled cavity and the reference frequency. When the two coupled cavities have the same orientation the value of the G-factor is maximised and the beat frequency is equal to $\omega_{+}\left(G=\text{max}\right)-\omega_{REF}$. After the second cavity has made a quarter rotation the geometry factor is equal to zero and the beat frequency is equal to $\omega_{+}\left(G=0\right)-\omega_{REF}$. Hence the fractional stability of the cavity / reference beat frequency for an integration time equal to a quarter of the rotation period is
\begin{align}
\delta\omega=\frac{\left(\omega_{+}\left(G=\text{max}\right)-\omega_{REF}\right)-\left(\omega_{+}\left(G=0\right)-\omega_{REF}\right)}{\omega_{0}} \nonumber \\
=\frac{\omega_{+}\left(G=\text{max}\right)-\omega_{+}\left(G=0\right)}{\omega_{0}}, \label{eq:deltaomega}
\end{align}
where $\omega_{+}$ is defined in equation~\eqref{eq:freqcoupled}. This is the stability of the frequency shift first illustrated in Fig.~\ref{fig:coupledfreqs} (red label) and then plotted explicitly in figure~\ref{fig:freqshift} as a function of frequency detuning and cavity losses. There are two regimes that the experiment could operate in. The first is the ideal situation where the cavities feature extremely high quality factors and suffer from minimal frequency detuning such that 
\begin{equation}
Q^{-1}+x<\chi^{2},
\label{eq:cond}
\end{equation}
in this scenario the frequency shift is independent of the detuning and cavity losses and is proportional to $\chi^{2}$. When the conditions set by eq.~\eqref{eq:cond} are broken the frequency shift changes as a function of the detuning and cavity losses and is proportional to $\chi^{4}$. 

To find analytical sensitivity expressions we take a series expansion of $\omega_{+}$ around $\chi=0$ to second order in $\chi$, substitute back in to equation~\eqref{eq:deltaomega} and solve for $\chi$. In the situation where eq.~\ref{eq:cond} holds true we can approximate the experimental sensitivity as
\begin{equation}
\chi\approx\sqrt{\frac{2\delta\omega\omega_{0}^{4}}{\sqrt{N G}\pmass^{4}}},
\label{eq:sensideal}
\end{equation}
where N is the number of measurements taken. This shows that at a fundamental level the frequency signal is proportional to $\chi^{2}$ whereas traditional LSW power measurements are proportional to $\chi^{4}$. Unfortunately for any experiment implemented using current technology it is likely that the conditions in eq.~\eqref{eq:cond} will be broken and the experimental sensitivity will be
\begin{equation}
\chi=\left(\frac{\sqrt{2}\delta\omega\zeta\omega_{0}^{8}}{\sqrt{N}G^{2}\pmass^{8}}\right)^{\frac{1}{4}}, \label{eq:chisens}
\end{equation}
where $\zeta$ is a term describing the quality factors and frequency detuning of the cavities,
\begin{equation}
\begin{split}
\zeta=\left(x^{2}-2\right)^{1.5}\left(\left(\frac{1}{Q_{1}Q_{2}}+x^{2}\right) \times \right. \\
\left. \left(4\sqrt{\frac{1}{Q_{1}Q_{2}}+x^{2}}-4-\frac{2}{Q_{1}Q_{2}}-x^{2}\right)\right)^{\frac{1}{2}}.
\end{split}
\end{equation}
One would expect that this $\zeta$ term will be the limiting factor of any experiment undertaken with current technology.

Using equation~\eqref{eq:sensideal} we can estimate bounds on $\chi$ that could be obtained with an ideal rotating cavity experiment. For a pair of cavities with an optimal fractional frequency stability of $3\times 10^{-16}$ that are operated for 1 week with a rotation period of 60 seconds, $\chi\approx2\times 10^{-9}$ assuming that $\pmass = \omega_{0}$ and $G=1$. If the experiment were to run for 9 weeks then $\chi$ could be bounded on the order of $10^{-10}$, which is at least two orders of magnitude lower than any existing experimental bounds for a paraphoton mass below 1~eV.

More conservative estimates for bounding $\chi$ can be made using eq.~\eqref{eq:chisens}. For a pair of cavities with a common central frequency of 10 GHz and a frequency detuning of 1 kHz an optimum fractional frequency stability of $10^{-14}$ is achievable with quality factors on the order of $10^{8}$~\cite{NiCav}. Setting the geometry factor to 1, assuming that $\pmass = \omega_{0}$ and running the experiment for 1 week with a rotation period of 60 seconds gives $\chi\approx3\times10^{-6}$. This limit is an improvement over existing bounds given by microwave cavity LSW power experiments in the same frequency/mass range~\cite{povey2010}, although bounds derived from indirect measurements such as cosmic microwave background data~\cite{cmb1,cmb2} and Coulomb law experiments~\cite{coulomb1,coulomb2} have set lower limits. If we assume this experiment can be run at the current state-of-the-art then $\chi$ bounds on the order of $10^{-7}$ could be obtained. Realistic improvements in technology would allow $\chi$ to be bounded on the order of $10^{-8}$.

\section{Conclusion}

We have derived the fundamental normal mode frequencies for a pair of resonant mode cavities coupled by the exchange of hidden sector photons and have demonstrated that frequency measurement experiments could be used to constrain the paraphoton kinetic mixing parameter, $\chi$. For the X-band microwave frequency region of hidden sector photon parameter space the sensitivity of our proposed experiment is comparable to current LSW power measurements. A rotating coupled cavity experiment would suppress systematic error due to power leakage which can be a technical limitation of LSW power experiments. The coupled cavity experiment developed here can be applied to other frequency ranges, including the optical domain, and to searches for any hypothetical particle that kinetically mixes with the photon.

\begin{acknowledgments}
We thank the Department of Energy's Institute for Nuclear Theory at the University of Washington for its hospitality and the Department of Energy for partial support during the completion of this work. This work was partially supported by the Australian Research Council Discovery Projects DP1092690 and DP130100205.
\end{acknowledgments}

\bibliography{biblo}

\end{document}